**Magneto-optical hologram lens with microsecond focal switching in magnetic garnet**

Hibiki Miyashita[1], Toshiaki Watanabe[2], Eri Negishi[2], Mitsuteru Inoue[1], Kazushi Ishiyama[1], Taichi Goto[1,*]

[1] Research Institute of Electrical Communication, Tohoku University, 2-1-1 Katahira, Aoba, Sendai, Miyagi 980-8577, Japan.

[2] Shin-Etsu Chemical Co., Ltd., 2-13-1 Isobe, Annaka, Gunma 379-0195, Japan.

**Contact Information**

* Corresponding author: Taichi Goto, (Dr., Prof.)

Phone: +81-22-217-5489

Email: taichi.goto.a6@tohoku.ac.jp

Address: Research Institute of Electrical Communication, Tohoku University, 2-1-1 Katahira, Aoba, Sendai, Miyagi 980-8577, Japan

**Abstract**

Active lenses based on metasurfaces have been demonstrated across broad applications, and their response speeds remain limited to the microsecond range. Here we report the experimental demonstration of a magneto-optical (MO) active lens based on a magnetic hologram formed on a magnetic garnet film. A Fresnel zone plate pattern was created using an array of permanent magnets, focusing a 633 nm laser beam of 12.5 mm diameter to a spot size of 0.7 mm at a focal length of 3.1 m with a modulation depth of ~91%. The focusing state was reversibly switched by applying an external magnetic field pulse. Through optimization of the magnetic field application system, a switching speed of 10.8 μs was achieved, surpassing the operating speeds of conventional active lenses. The intrinsic picosecond-scale response of the MO effect indicates that further improvement in switching speed is achievable, establishing this work as a proof-of-concept for ultra-fast active lenses.

**Keywords**

Magnetooptical effect, Magnetic domain, Fresnel lens, Magnetic hologram, Iron Garnet, Magnetic pulse field

## 1. Introduction

The control speed and dynamic range of focal points in traditional lenses are fundamentally limited by the mechanical movement of optical elements. To overcome these limitations, metalenses formed from artificially engineered nano- and microscale structures known as metasurfaces have attracted significant attention across the fields of nanomaterials, optics, and electronics. As reviewed by Abdelraouf et al.[1], tunable metasurfaces cover a wide range of operating wavelengths from the visible[2] to the terahertz regime[3], and numerous modulation schemes have been proposed, including ferroelectric[4] and liquid crystal approaches[5]. For instance, Mukherjee et al. demonstrated focal switching of a lens using ferroelectric liquid crystals[4]. Onaka et al. realized an active lens by controlling liquid crystal orientation with piezoelectrically generated ultrasound[5]. Schneider et al. demonstrated a tunable focal length using a piezoelectrically driven silicon-membrane lens[6,7]. Such active lenses have been demonstrated across a broad range of applications, including microscopic and biomedical imaging[8-14], laser processing and trapping[15,16], medical optics[17-20], and augmented-reality devices[21,22]. However, their response speed is typically in the microsecond range[23], leaving them insufficient for demanding high-speed applications such as high-speed communication systems.

To overcome these speed constraints, the MO effect, particularly as exhibited by magnetic garnet films, presents a compelling physical mechanism for high-speed light modulation[24-29]. Prior research has demonstrated MO responses on the order of 100 fs in magnetic garnet films[30], and a film of ~100 μm thickness achieves 90° modulation, corresponding to a 100% modulation depth[27], exceeding that of electro-optic materials. Lens modulation exploiting the MO effect has been theoretically proposed[31], but no experimental demonstration has been reported.

Magnetic domains in MO materials can be patterned to form holographic structures functioning as metasurfaces[32,33], and these patterns are controllable by an external magnetic field. This domain-based approach offers several advantages over alternative metasurface designs. Unlike metasurfaces based on refractive-index modulation or nanoscale surface-relief structures[2,3], MO hologram metasurfaces introduce no additional light scattering. Unlike liquid crystal metasurfaces[4,17], no electrodes are required on the reflective surface—the field-applying coil can be placed on the reverse side of the device—yielding a completely flat metalens architecture free from spurious scattered light.

MO hologram recording has been studied extensively. Fan et al. reported an experimental demonstration of an MO hologram in 1969[34], followed by substantial improvements in reconstruction fidelity by Shirakashi and co-workers[35]. Hologram patterning by laser writing was further demonstrated by Fujita and co-workers[33]. However, active control of MO hologram patterns by external magnetic field pulses has not been pursued. The key obstacles have been the difficulty of fabricating electromagnets and drive circuits capable of delivering short, high-field pulses, and the challenge of producing magnetic garnet films exhibiting a large MO effect, perpendicular magnetic anisotropy, and a low saturation field over a large area. In the present work, we resolve both challenges by applying the high-field short-pulse circuit technology and MO garnet fabrication expertise established in our previous work on MO Q-switched lasers[27], and demonstrate that a metalens realized through an MO hologram pattern can be repeatedly and rapidly switched by an external magnetic field.

## 2. Design of MO Hologram Lens

To apply the MO effect to an active lens, we propose a diffractive lens based on a Fresnel Zone Plate (FZP) pattern formed by magnetic domains. The FZP consists of concentric annular rings arranged to shift the phase of transmitted light alternately by $+\pi/2$ and $-\pi/2$. The radius of the $n$th ring boundary $r_n$ is given by[36]

$$r_n = \sqrt{n\lambda F} \qquad (1)$$

where $\lambda$ is the wavelength of the incident light, $F$ is the designed focal length, and $n$ is a natural number ($n$ = 1, 2, 3, …). The focusing performance of this domain-based FZP was evaluated using the angular spectrum method[37]; details of the calculation procedure are provided in the Supplementary Information. Fig. 1(a) shows the ideal FZP domain distribution with a designed focal length $F = 4$ m, yielding a center circular domain of radius $r_1$ ~3.18 mm and a polarization rotation (PR) angle of $\pm\pi/4$, a typical value for magnetic garnet films. The simulation confirms focusing at $z = 4$ m with a peak intensity 358 times that of the incident beam and a beam diameter of 24.1 μm (Fig. 1b–d).

For practical fabrication, the ideal FZP pattern was simplified in two steps, described in detail in the Supplementary Information. The final simplified pattern consists of one central disk of diameter 2 mm and 174 surrounding disks of diameter 0.8 mm arranged into five rings (Fig. 1e), matching commercially available permanent magnet dimensions. Despite the simplification, the beam was successfully focused at $z = 3.35$ m with a spot size of 0.99 mm and an intensity enhancement of 11.4, corresponding to ~1/31 of the ideal FZP performance (Fig. 1f–h). This reduction is primarily attributable to the pattern simplification, and the result nonetheless confirms the feasibility of the proposed MO hologram lens concept.

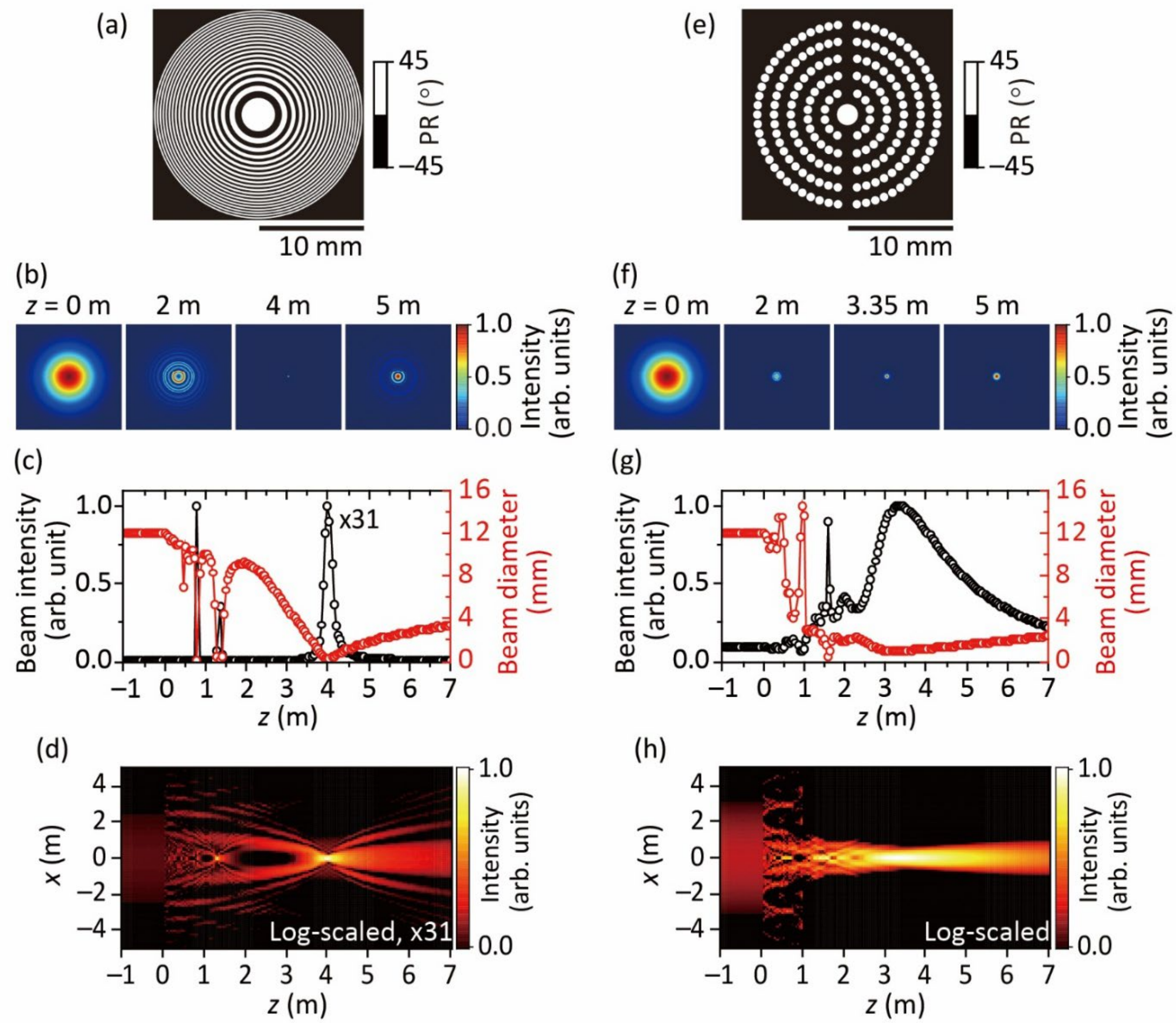

**FIGURE 1.** Diffraction simulation results using the angular spectrum method. (a, e) PR angle distributions at $z$ = 0 m: (a) ideal FZP pattern, and (e) simplified pattern using commercially available magnet dimensions. (b, f) Normalized Gaussian beam propagation profiles. (c, g) Beam diameter and normalized intensity as a function of propagation distance. (d, h) Normalized color maps of the modulated beam.

### 3. Device Structure and Fabrication

The MO hologram lens consists of a magnetic garnet film, a permanent magnet module, and a Helmholtz coil (Fig. 2a). The garnet film is coated with a high-reflection (HR) coating on the front surface and an anti-reflection (AR) coating on the substrate side (Fig. 2b), with details of the coating design provided in the Methods section. The permanent magnet module was constructed by placing $Sm_2Co_{17}$ magnets of diameter 0.8 mm in a brass plate with holes at positions corresponding to the simplified FZP pattern. A paper spacer was inserted between the permanent magnet module and the garnet film to reduce the external magnetic field required for switching. The magnets were bonded with a cyanoacrylate-based adhesive, as shown in the enlarged view of the module (Fig. 2f) and the photograph showing the overall appearance of the fabricated module (Fig. S3).

The magnetic hysteresis loop of the $Sm_2Co_{17}$ magnets, measured with a vibrating sample magnetometer (VSM), is shown in Fig. 2c. The saturation magnetization $M_{s,mag}$ was 707 kA/m and the residual magnetic flux density $B_{r,mag}$ was 851 mT, consistent with previously reported values for $Sm_2Co_{17}$ [38], confirming that the magnets generate a sufficient field to saturate the garnet film. The PR hysteresis loop of the garnet film at $\lambda = 660$ nm is shown in Fig. 2d, yielding a PR angle of $24.0 \pm 0.02°$ and saturation magnetic fields $H_{s,PR}$ of 11.2–19.3 mT. The in-plane (IP) and out-of-plane (OP) magnetic hysteresis loops of the garnet film are shown in Fig. 2e, from which the OP saturation field $H_{s,OP}$ of 13.9 ± 0.5 mT was obtained, with further magnetic characterization data provided in the Supplementary Information.

The calculated magnetic field distribution generated by the magnet module on the garnet film surface is shown in Fig. 2g, with details of the simulation conditions provided in the Supplementary Information. The magnetic field distribution during switching was further analyzed by simulation, confirming that an external field of $\mu_0 H_{ex}$ = 53 mT is sufficient to saturate the garnet film over the entire lens aperture, as described in the Supplementary Information (Fig. S2). The polarized-light microscope image of the garnet film placed on the magnet module (Fig. 2h), taken at λ = 470 nm, confirms the formation of a domain pattern consistent with the calculated field distribution. The central circular domain had a diameter of 4.58 ± 0.713 mm, and the width of the first annular domain was 0.947 ± 0.069 mm, demonstrating that the magnet arrangement successfully generated the intended FZP-like domain pattern.

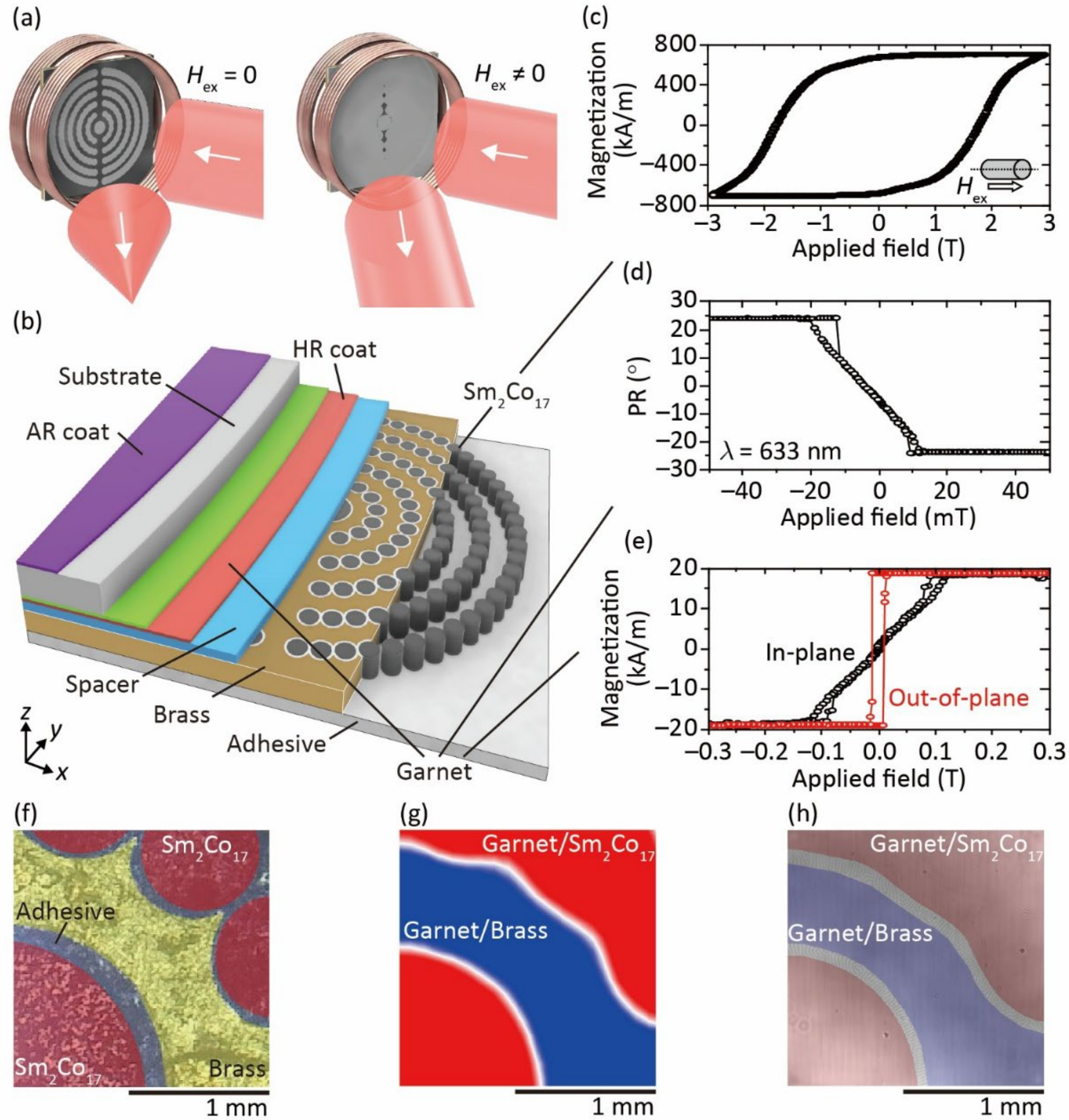

**FIGURE 2.** Device configuration of the MO hologram lens. (a) Schematic of the device, where an external magnetic field $H_{ex}$ switches the focusing state. (b) Layer structure of the device. (c) Longitudinal magnetic hysteresis loop of the $Sm_2Co_{17}$ magnet rod measured by VSM. (d) PR hysteresis loop of the magnetic garnet film at $\lambda = 660$ nm. (e) IP and OP magnetic hysteresis loops of the magnetic garnet film. (f) Enlarged view of the permanent magnet module. (g) Calculated magnetic field distribution on the garnet film surface generated by the magnet module. (h) Polarized-light microscope image of the garnet film placed on the magnet module at $\lambda = 470$ nm.

## 4. Demonstration of Focal Switching

To demonstrate the functionality of the MO hologram lens, we constructed the optical system shown in Fig. 3a, with details of the optical components and measurement setup provided in the Methods section. A He-Ne laser at $\lambda$ = 633 nm was expanded to a beam diameter of 12.5 mm and directed onto the MO hologram lens at an angle of 3°.

The device successfully focused the reflected beam at $z$ = 3.1 m (Fig. 3b), slightly shorter than the theoretically predicted focal length of 3.35 m. This discrepancy is attributed to the actual magnetic domain pattern deviating from the ideal FZP shape, as confirmed by the field distribution shown in Fig. S2. The power density increased from 7.3 $W/m^2$ at $z$ = 0.9 m to 22.2 $W/m^2$ at the focal point ($z$ = 3.1 m), while the beam diameter was reduced from 2.1 mm to 0.7 mm, corresponding to a power density enhancement of ~3.0 times and a beam diameter reduction to ~33% of the incident size. The beam profile at the focal point showed a well-defined single spot, whereas annular profiles were observed before and after the focal point (Fig. 3c). When an external field of $\mu_0 H_{ex}$ = 53 mT was applied, the power density remained nearly constant across the entire measured range, reaching only 1.7 $W/m^2$ at $z$ = 3.1 m, corresponding to ~7.7% of the value at 0 mT and a switching contrast of ~13.1, consistent with the simulated value of 11.4. This focusing performance remains lower than that reported for liquid crystal FZP lenses, where a power density enhancement of 420 times the incident intensity has been achieved[39], and thus improvement is expected by bringing the hologram pattern closer to the ideal FZP shape. The depth of focus was 1.59 m, sharper than the calculated value of 2.49 m, attributed to the adjacent disk domains merging into a continuous ring shape, bringing the pattern closer to the ideal FZP structure. A residual light intensity distribution along the *y*-axis was observed at $z$ = 0.9 m under $\mu_0 H_{ex}$ = 53 mT, attributed to the incomplete saturation of the garnet film near the center of the brass fixture, as shown in Fig. S2c.

The focusing state was further confirmed to be reversibly switchable by applying magnetic field pulse trains of amplitude 53 mT. The reflected light power at the focal point $z$ = 3.1 m was modulated between 1.6 µW at $\mu_0 H_{ex}$ = 0 mT and 17.3 µW at $\mu_0 H_{ex}$ = 53 mT, corresponding to a modulation factor of ~10.8 and a modulation depth of ~91%, with nearly identical beam profiles reproduced under repeated pulse application (Fig. 3d). However, the switching speed was 7.2 ms, limited by the slow rise time of the applied magnetic field pulse, attributable to the use of an oversized Helmholtz coil of diameter 14 cm and a power supply not optimized for high-frequency operation. The approach to achieving faster switching is described in Section 5.

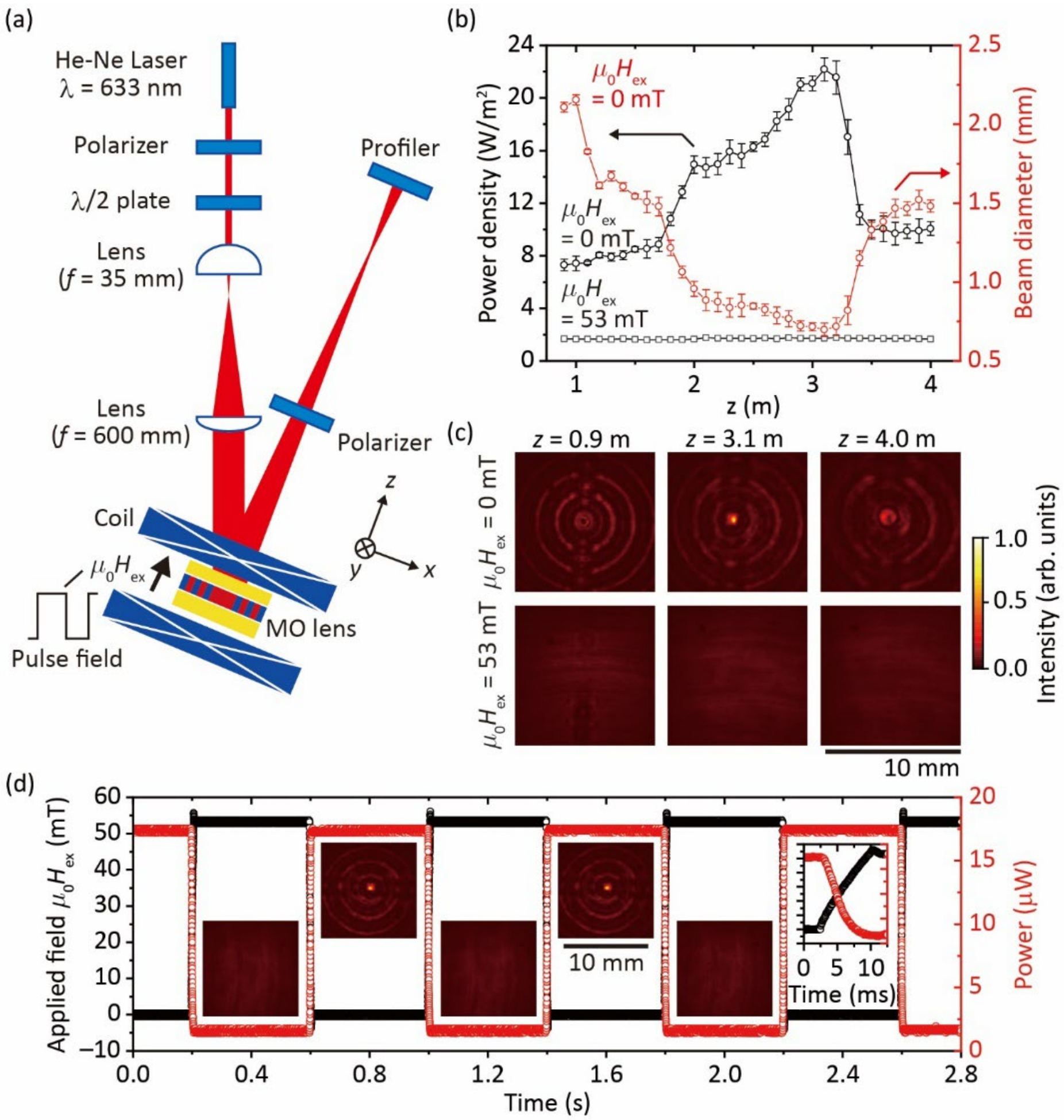

**FIGURE 3.** Demonstration of MO holographic lens controlled by magnetic field. (a) Optical setup used in evaluation of MO lens. (b) Power density and beam diameter of the laser modulated by MO lens. (c) Beam profiles at distances *z* of 0.9 m, 3.1 m, and 4.0 m for both fields of 0 mT and 53 mT. (d) Focal point's power modulated magnetic pulse trains with an amplitude of 53 mT at *z* = 3.1 m. Insets show the beam profiles for each timing.

**5. Microsecond Focal Switching**

The 7.2 ms switching speed demonstrated in Section 4 is limited not by the intrinsic response of the magnetic domains but by the slow rise time of the applied magnetic field pulse, attributable to the oversized Helmholtz coil and the low-frequency power supply. To overcome this limitation, we replaced the Helmholtz coil with a compact solenoid coil matched to the beam diameter and drove it with a high-frequency pulse circuit, following the approach established in our previous work on MO Q-switched lasers[27], with details of the circuit and coil parameters provided in the Methods section. The solenoid coil was fabricated with 5.5 turns of 0.5 mm diameter enameled copper wire with an inner diameter of 12.5 mm, matching the laser beam diameter. Since the magnetic field generated by the coil alone was insufficient to switch the MO hologram lens, a neodymium magnet of 20 mm diameter and 10 mm thickness was positioned 18 mm from the brass fixture side of the device to provide a bias field. The coil was fixed to the AR-coated surface of the device, as shown in Fig. 4a.

The resulting magnetic field pulse exhibited a full width at half maximum (FWHM) of ~490 ns (Fig. 4b, c). In response to this pulse, the focal point intensity at $z$ = 3.1 m was modulated with a switching speed of 10.8 μs (Fig. 4d, e), representing an improvement of more than two orders of magnitude over the 7.2 ms achieved in Section 4 and surpassing the operating speeds of conventional active lenses based on ferroelectric liquid crystals[4]. The reflected light power was modulated with a modulation depth of ~10% through magnetic field switching, which is lower than the ~91% achieved in Section 4 due to insufficient magnetic field strength resulting from the solenoid coil not being fully optimized for impedance matching and high-frequency operation. Nearly identical power modulation was confirmed under repeated pulse application. A discrepancy is observed between the shapes of the magnetic field pulse (Fig. 4c) and the optical response (Fig. 4e), attributed to the limited frequency response of the photodetector, which was configured to prioritize measurement sensitivity. The actual optical response is therefore expected to follow the ~490 ns FWHM magnetic field pulse more closely than the measured 10.8 μs.

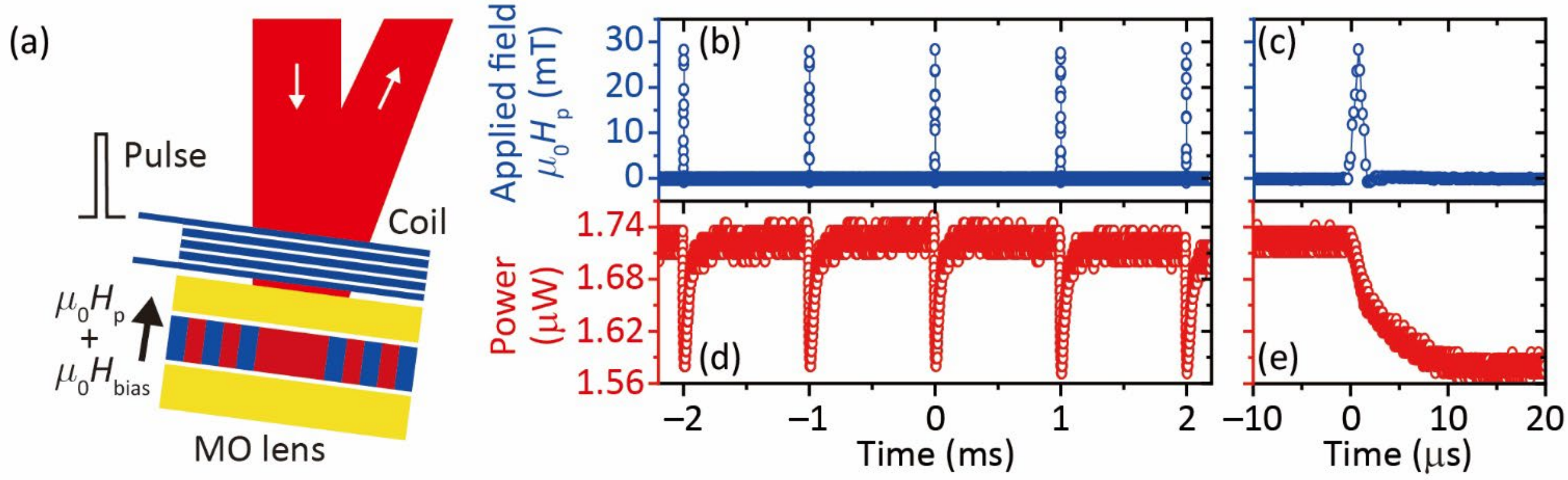

**FIGURE 4.** Microsecond focal switching of the MO hologram lens modulated by a short magnetic field pulse. (a) Arrangement of the solenoid coil and neodymium magnet relative to the MO hologram lens. (b, c) Magnetic field pulses generated by the coil at different time scales. (d, e) Focal point power modulation at $z$ = 3.1 m in response to the magnetic field pulses shown in (b) and (c), respectively.

## 6. Conclusion

This work demonstrates a magnetic hologram active lens based on the MO effect. A laser beam with a wavelength of 633 nm and a diameter of 12.5 mm was focused to a spot diameter of 0.7 mm at a focal length of 3.1 m with a modulation depth of ~91%, consistent with the theoretical prediction. High-speed switching of the focusing state was demonstrated by applying short magnetic field pulses to the magnetic garnet film. Through optimization of the magnetic field application system, a switching speed of 10.8 μs was achieved. The focal length of 3.1 m is determined by the minimum magnet feature size of 0.8 mm, and micropatterning or writing of the magnetic hologram structure would enable both shorter focal lengths and improved focusing performance. The switching speed of 10.8 μs demonstrated here surpasses those of conventional active lenses based on liquid crystal and membrane technologies[23]. The intrinsic MO response of magnetic garnet films has been demonstrated to reach the femtosecond regime[30], indicating that the switching speed of MO hologram lenses is not fundamentally limited by the material but by the magnetic field application system. Further reduction of the required switching field will enable the use of faster driving circuits without a bias magnet, providing a pathway toward active lenses operating in the nanosecond or even subnanosecond range.

## Method

***Sample Deposition.*** The HR and AR coatings were deposited using a radio frequency ion beam sputtering (RF-IBS) system (RMTec, RM17-0010) with a base pressure of $1 \times 10^{-4}$ Pa. During deposition, 8 sccm of $O_2$ was introduced at a working pressure of $1 \times 10^{-1}$ Pa. The substrate was rotated at 10 rpm to ensure film uniformity. 4-inch diameter sintered $SiO_2$ and $Ta_2O_5$ targets (Furuuchi Kagaku Co.) were used. The ion beam voltage and current were 800 V and 36 mA, respectively, with 10 and 14 sccm of Ar flowing to the ion gun and low-frequency neutralizer (LFN), respectively. The ion beam accelerator voltage was 160 V and the RF power was 60 W. The deposition rates were 2.05 nm/min for $SiO_2$ and 1.03 nm/min for $Ta_2O_5$. The HR coating consisted of a $(SiO_2/Ta_2O_5)^5$ multilayer stack with layer thicknesses of 108 nm for $SiO_2$ and 71 nm for $Ta_2O_5$, yielding a theoretical reflectance of 93.8% at $\lambda$ = 633 nm. The AR coating was a $SiO_2/Ta_2O_5/SiO_2$ multilayer deposited on the substrate with thicknesses of 300 nm, 103 nm, and 25 nm, respectively, yielding a theoretical transmittance of 99.7% at $\lambda$ = 633 nm.

***MO characterization.*** MO properties were characterized using an MO measurement system (JASCO, J-1700FK). This system is based on a rotating polarizer and polarization modulation methods. The light was incident on a garnet film at an angle of 5°, and the PR angle of the reflected light was measured. The garnet film temperature was maintained at 40±0.5°C using a Peltier temperature controller to reduce the effects of the temperature dependence of garnet magnetization. A magnetic field of ±50 mT was applied.

***Magnetic properties characterization.*** Magnetic properties were measured with a vibrating sample magnetometer (VSM, Tamakawa Corp., TM-VSM2050-HGC-TOH), and an ionizing fan (AS ONE Corp., SIB-1DC) was used to suppress the electrostatic noise of the VSM. The measurement was performed on an uncoated sample at room temperature, and the paramagnetic background was subtracted. A magnetic field of ±1.2 T was applied.

***Simulation of magnetic field in MO lens.*** The magnetic field distribution in the MO hologram lens was simulated using a three-dimensional finite integration technique (FIT) based simulator (CST Studio Suite 2022 SP5). The permanent magnet module was modeled with the same geometry as in Fig. 2f. The space between the magnets and the brass plate was filled with adhesive models. The MO lens was modeled as a seven-layer structure consisting of an AR coating, a substrate, the magnetic garnet film, an HR coating, a paper spacer, the permanent magnet module, and the cured adhesive, with material parameters and layer thicknesses summarized in Table S1. The AR and HR coatings were simplified as single-layer film models. The magnetic permeability of the garnet film was set based on the OP measurement results from

Fig. 2e. The magnetization characteristics of the permanent magnets were set to the manufacturer's nominal properties, with a residual magnetic flux density of 1.06 T.

***Optical setup used in evaluation of MO lens.***

The MO lens was fixed inside a Helmholtz coil, and the magnetic domain pattern was modulated by the magnetic field generated by the coil. A square wave signal was generated by a signal generator (Tektronix, AFG3022), amplified by a bipolar power supply (TAKASAGO, BWS 40-15), and used to drive a pulsed current through the coil. The coil current was measured with a current probe (Tektronix, TCP312) and an oscilloscope (Keysight, InfiniiVision DSOX3034G) to control the magnetic field generated by the coil to be $\mu_0 H_{ex} = 53$ mT. The reflected laser light profile was measured using a laser beam profiler (GENTEC-EO, BEAMAGE-4M) at 0.1 m intervals from $z = 0.9$ m to 4.0 m, with the MO lens surface defined as $z = 0.0$ m. The profiler's capture timing was controlled by a trigger signal from the signal generator. The modulation of the light intensity was measured using a photodetector (Thorlabs, DET10A/M), and the optical power was measured using a power meter (Newport, 818-SL).

***Quick modulation setup of MO lens.***

The short-pulse magnetic field was generated using a circuit following the same schematic described in Ref. [27], with modified component values of $R_2$ = 4.99 Ω, $R_3$ = 0 Ω, $R_4$ = 1.2 kΩ, $C_1$ = 10 μF, and $C_2$ = 0.47 μF. The solenoid coil had a resistance of $R_1$ = 37.4 mΩ and an inductance of $L$ = 490 nH. The termination resistance of the oscilloscope connected to the photodetector was set to 100 kΩ, balancing signal intensity and the frequency response of the detector.

**Conflict of Interest**

The authors declare no conflict of interest.

**Acknowledgements**

This work was partly supported by JSPS KAKENHI (Nos. 23H01439, and 23K17758), NEDO (No. 23200047), the Amada Foundation. T.G. acknowledge receiving support by TI-FRIS.

**REFERENCES**

1 Abdelraouf, O. A. M. *et al.* Recent Advances in Tunable Metasurfaces: Materials, Design, and Applications. *ACS Nano* **16**, 13339-13369, doi:10.1021/acsnano.2c04628 (2022).

2 Huang, P.-S. *et al.* Varifocal Metalenses: Harnessing Polarization-Dependent Superposition for Continuous Focal Length Control. *Nano Lett.* **23**, 10432-10440, doi:10.1021/acs.nanolett.3c03056 (2023).

3 Yang, W. *et al.* A self-biased non-reciprocal magnetic metasurface for bidirectional phase modulation. *Nature Electronics* **6**, 225-234, doi:10.1038/s41928-023-00936-w (2023).

4 Mukherjee, S. *et al.* Fast refocusing lens based on ferroelectric liquid crystals. *Opt. Express* **29**, 8258-8267, doi:10.1364/OE.417112 (2021).

5 Onaka, J., Iwase, T., Fukui, M., Koyama, D. & Matsukawa, M. Ultrasound liquid crystal ens with enlarged aperture using raveling waves. *Opt. Lett.* **46**, 1169-1172, doi:10.1364/Ol.414295 (2021).

6 Schneider, F., Draheim, J., Kamberger, R., Waibel, P. & Wallrabe, U. Optical characterization of adaptive fluidic silicone-membrane lenses. *Opt. Express* **17**, 11813-11821, doi:10.1364/Oe.17.011813 (2009).

7 Schneider, F., Draheim, J., Müller, C. & Wallrabe, U. Optimization of an adaptive PDMS-membrane lens with an integrated actuator. *Sens. Actuators, A: Physical* **154**, 316-321, doi:10.1016/j.sna.2008.07.006 (2009).

8 Kaplan, A., Friedman, N. & Davidson, N. Acousto-optic lens with very fast focus scanning. *Opt. Lett.* **26**, 1078-1080, doi:10.1364/OL.26.001078 (2001).

9 Yang, L., Mac Raighne, A., McCabe, E. M., Dunbar, L. A. & Scharf, T. Confocal microscopy using variable-focal-length microlenses and an optical fiber bundle. *Appl. Opt.* **44**, 5928-5936,

doi:10.1364/AO.44.005928 (2005).

10 Yang, V. X. D. *et al.* Doppler optical coherence tomography with a micro-electro-mechanical membrane mirror for high-speed dynamic focus tracking. *Opt. Lett.* **31**, 1262-1264, doi:10.1364/OL.31.001262 (2006).

11 Sancataldo, G. *et al.* Three-dimensional multiple-particle tracking with nanometric precision over tunable axial ranges. *Optica* **4**, 367-373, doi:10.1364/OPTICA.4.000367 (2017).

12 Fahrbach, F. O., Voigt, F. F., Schmid, B., Helmchen, F. & Huisken, J. Rapid 3D light-sheet microscopy with a tunable lens. *Opt. Express* **21**, 21010-21026, doi:10.1364/OE.21.021010 (2013).

13 Jiang, J. *et al.* Fast 3-D temporal focusing microscopy using an electrically tunable lens. *Opt. Express* **23**, 24362-24368, doi:10.1364/OE.23.024362 (2015).

14 Asatryan, K. *et al.* Optical lens with electrically variable focus using an optically hidden dielectric structure. *Opt. Express* **18**, 13981-13992, doi:10.1364/OE.18.013981 (2010).

15 Kuwano, R., Tokunaga, T., Otani, Y. & Umeda, N. Liquid Pressure Varifocus Lens. *Optical Review* **12**, 405-408, doi:10.1007/s10043-005-0405-3 (2005).

16 Kawamura, M., Ye, M. & Sato, S. Optical Trapping and Manipulation System Using Liquid-Crystal Lens with Focusing and Deflection Properties. *Jpn. J. Appl. Phys.* **44**, 6098, doi:10.1143/jjap.44.6098 (2005).

17 Chen, H.-S. & Lin, Y.-H. An endoscopic system adopting a liquid crystal lens with an electrically tunable depth-of-field. *Opt. Express* **21**, 18079-18088, doi:10.1364/OE.21.018079 (2013).

18 Li, G. *et al.* Switchable electro-optic diffractive lens with high efficiency for ophthalmic applications. *Proceedings of the National Academy of Sciences* **103**, 6100-6104, doi:10.1073/pnas.0600850103 (2006).

19 Lin, Y.-H. & Chen, H.-S. Electrically tunable-focusing and polarizer-free liquid crystal lenses for ophthalmic applications. *Opt. Express* **21**, 9428-9436, doi:10.1364/OE.21.009428 (2013).

20 Li, G. *et al.* High-efficiency switchable flat diffractive ophthalmic lens with three-layer electrode pattern and two-layer via structures. *Appl. Phys. Lett.* **90**, 111105, doi:10.1063/1.2712773 (2007).

21 Kumar, M. B. *et al.* Compact vari-focal augmented reality display based on ultrathin, polarization-insensitive, and adaptive liquid crystal lens. *Optics and Lasers in Engineering* **128**, 106006, doi:10.1016/j.optlaseng.2020.106006 (2020).

22 Shin, D., Kim, C., Koo, G. & Hyub Won, Y. Depth plane adaptive integral imaging system using a vari-focal liquid lens array for realizing augmented reality. *Opt. Express* **28**, 5602-5616, doi:10.1364/OE.384697 (2020).

23 Kang, S., Duocastella, M. & Arnold, C. B. Variable optical elements for fast focus control. *Nat. Photon.* **14**, 533-542, doi:10.1038/s41566-020-0684-z (2020).

24 Stupakiewicz, A., Szerenos, K., Afanasiev, D., Kirilyuk, A. & Kimel, A. V. Ultrafast nonthermal photo-magnetic recording in a transparent medium. *Nature* **542**, 71-74, doi:10.1038/nature20807 (2017).

25 Avci, C. O. *et al.* Fast switching and signature of efficient domain wall motion driven by spin-orbit torques in a perpendicular anisotropy magnetic insulator/Pt bilayer. *Appl. Phys. Lett.* **111**, 072406, doi:10.1063/1.4994050 (2017).

26 Yajima, S., Nishiyama, N. & Shoji, Y. High-speed modulation in a waveguide magneto-optical switch with impedance-matching electrode. *Opt. Express* **31**, 16243-16250, doi:10.1364/OE.480835 (2023).

27 Goto, T. *et al.* Magneto-optical Q-switching using magnetic garnet film with micromagnetic domains. *Opt. Express* **24**, 17635-17643, doi:10.1364/OE.24.017635 (2016).

28 Morimoto, R. *et al.* Magnetic domains driving a Q-switched laser. *Sci. Rep.* **6**, 38679, doi:10.1038/srep38679 (2016).

29 Morimoto, R. *et al.* Randomly polarised beam produced by magnetooptically Q-switched laser. *Sci. Rep.* **7**, 15398, doi:10.1038/s41598-017-15826-3 (2017).

30 Kimel, A. V. *et al.* Ultrafast non-thermal control of magnetization by instantaneous photomagnetic pulses. *Nature* **435**, 655-657, doi:10.1038/nature03564 (2005).

31 Shamuilov, G., Domina, K., Khardikov, V., Nikitin, A. Y. & Goryashko, V. Optical magnetic lens: towards actively tunable terahertz optics. *Nanoscale* **13**, 108-116, doi:10.1039/D0NR06198K (2021).

32 Nakamura, Y., Lim, P. B., Goto, T., Uchida, H. & Inoue, M. Recording and reconstruction of volumetric magnetic hologram using multilayer medium with heat dissipation layers. *Opt. Express* **27**, 27573-27579, doi:10.1364/OE.27.027573 (2019).

33 Fujita, T. *et al.* Magneto-optical diffractive deep neural network. *Opt. Express* **30**, 36889-36899, doi:10.1364/OE.470513 (2022).

34 Fan, G., Pennington, K. & Greiner, J. H. Magneto-Optic Hologram. *J. Appl. Phys.* **40**, 974-975, doi:10.1063/1.1657805 (1969).

35 Shirakashi, Z. *et al.* Reconstruction of non-error magnetic hologram data by magnetic assist recording. *Sci. Rep.* **7**, 12835, doi:10.1038/s41598-017-12442-z (2017).

36 Kirz, J. Phase Zone Plates for X-Rays and Extreme Uv. *J. Opt. Soc. Am.* **64**, 301-309, doi:10.1364/Josa.64.000301 (1974).

37 Goodman, J. W. *Introduction to Fourier Optics*. (The McGRAW-HILL, 1996).

38 Sun, W. *et al.* Magnetic properties and microstructures of high-performance $Sm_2Co_{17}$ based alloy. *J. Magn. Magn. Mater.* **378**, 214-216, doi:10.1016/j.jmmm.2014.11.026 (2015).

39 Mel'Nikova, E. A. *et al.* Electrically Controlled Liquid Crystal Twist-Planar Fresnel Lens.

*Journal of Applied Spectroscopy* **91**, 1072-1077, doi:10.1007/s10812-024-01822-9 (2024).

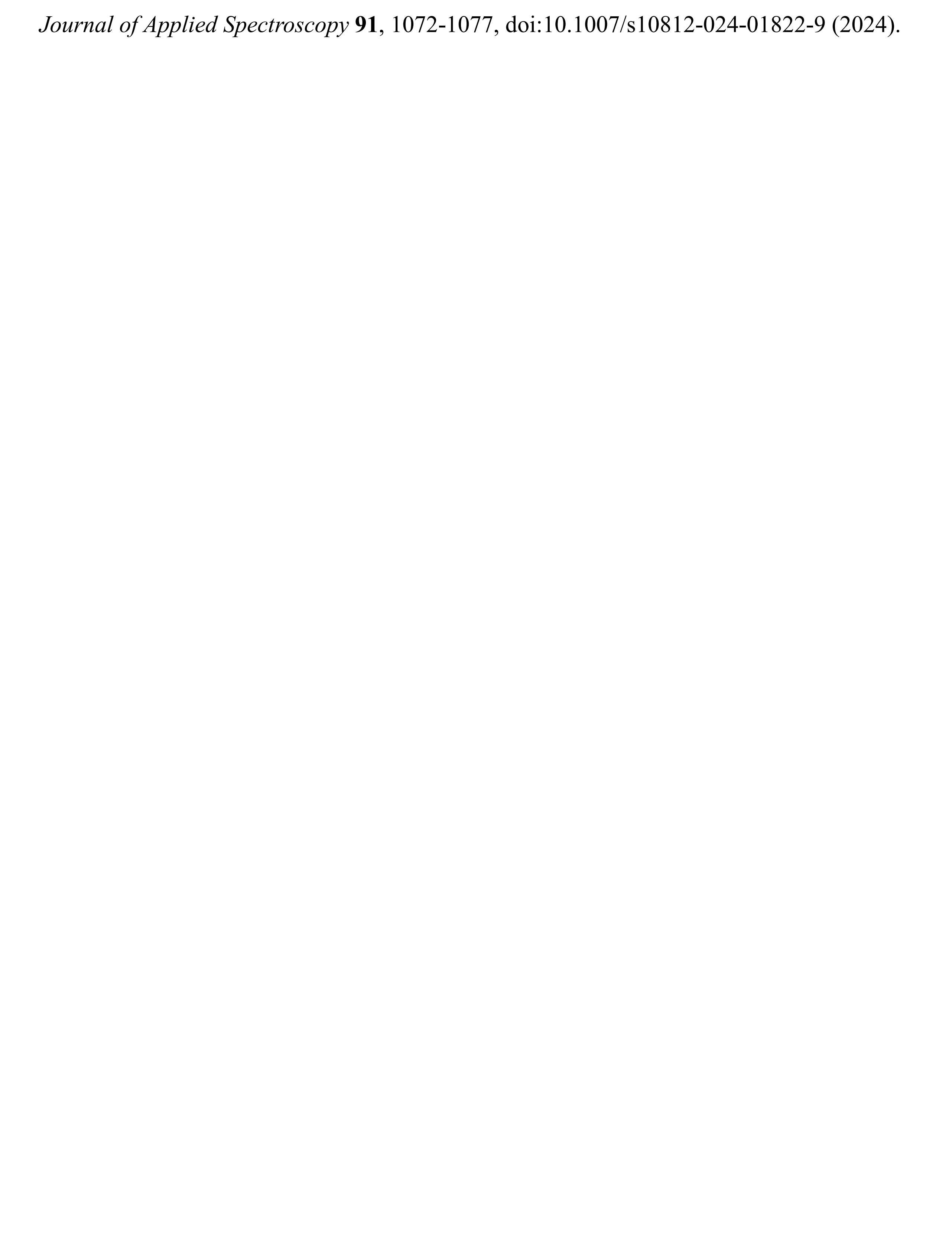

**Supplementary Information**

# Magneto-optical hologram lens with microsecond focal switching in magnetic garnet

Hibiki Miyashita[1], Toshiaki Watanabe[2], Eri Negishi[2], Mitsuteru Inoue[1], Kazushi Ishiyama[1], Taichi Goto[1,*]

[1] Research Institute of Electrical Communication, Tohoku University, 2-1-1 Katahira, Aoba, Sendai, Miyagi 980-8577, Japan.

[2] Shin-Etsu Chemical Co., Ltd., 2-13-1 Isobe, Annaka, Gunma 379-0195, Japan.

**Contact Information**

* Corresponding author: Taichi Goto, (Dr., Prof.)

Phone: +81-22-217-5489

Email: taichi.goto.a6@tohoku.ac.jp

Address: Research Institute of Electrical Communication, Tohoku University, 2-1-1 Katahira, Aoba, Sendai, Miyagi 980-8577, Japan

**Contents of Supplementary Information**

## Supplementary Note 1: Angular Spectrum Method

The light propagation calculations were performed using the angular spectrum method (ASM), a Fourier transform-based technique for solving the scalar diffraction problem. A light source at $z = -1$ m generated a Gaussian beam with a wavelength of $\lambda = 633$ nm and a beam diameter of 12.5 mm. The polarization rotation (PR) angle distribution was applied at the $z = 0$ m plane. The calculation range extended from z = –1 m to $z = 7$ m in steps of 0.05 m.

## Supplementary Note 2: Simplification of FZP Pattern

The ideal FZP pattern was simplified in two steps to enable fabrication using commercially available permanent magnets (Fig. S1). First, the ideal pattern consisting of 1 central disk and 19 rings of varying width (Fig. 1a) was modified to consist of 1 central disk and 5 rings of equal width (Fig. S1a), based on the principle that the inner rings contribute more significantly to focusing than the outer rings. Second, each ring was approximated by arranging small disks of diameter 0.8 mm in a ring shape (Fig. S1c). To prevent the gaps between rings from becoming entirely discrete, two small disks were removed from each ring. The resulting simplified pattern consists of one central disk of diameter 2 mm and 174 surrounding disks of diameter 0.8 mm arranged into five rings (Fig. S1e). The beam propagation profiles corresponding to each simplification step are shown in Figs. S1b, d, and f. The simplified pattern achieves a focal intensity of ~1/31 of the ideal FZP, primarily due to the extensive pattern simplification.

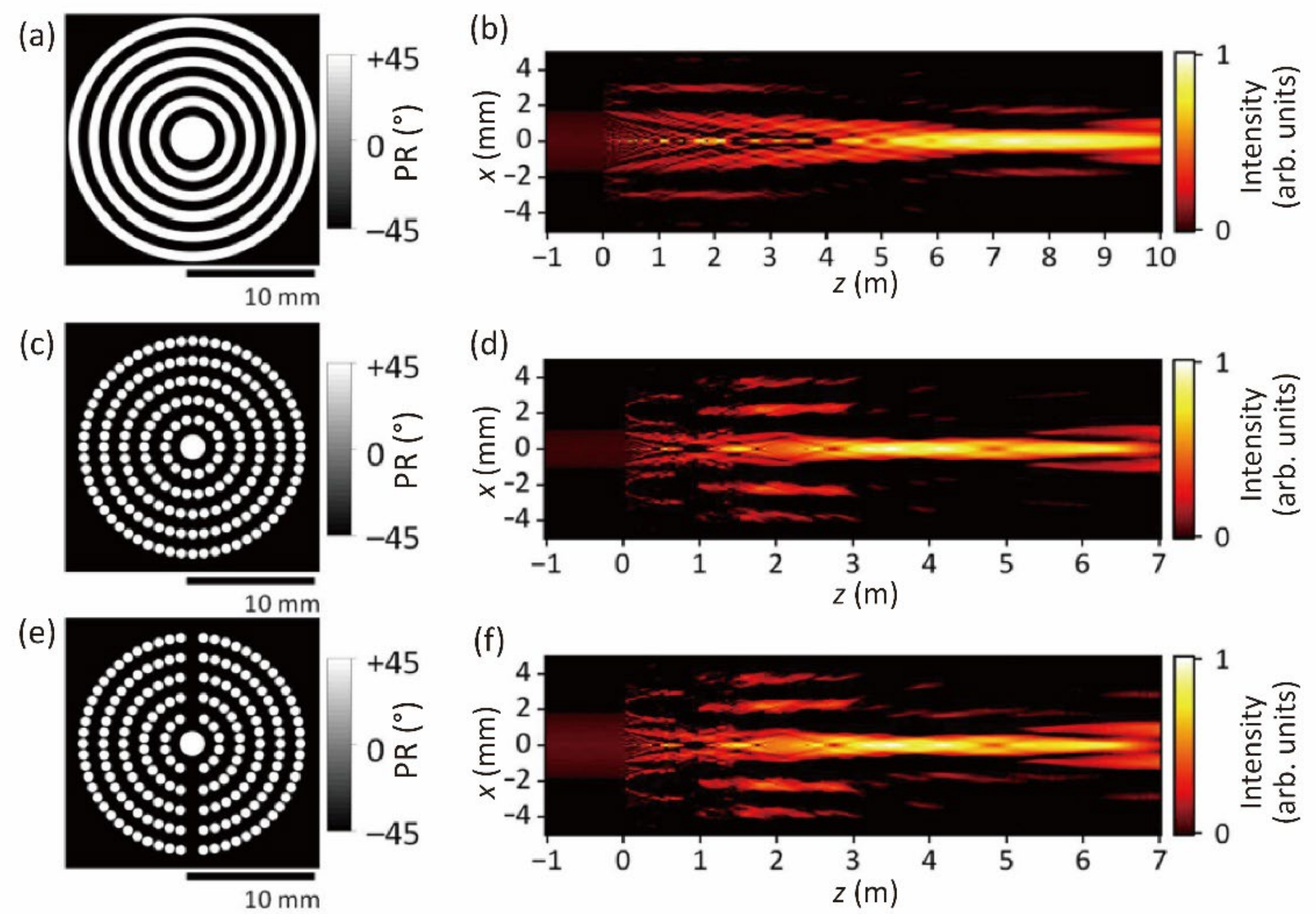

**FIGURE S1.** Simplification process of the FZP domain pattern and corresponding beam propagation simulations. (a, c, e) Polarization rotation (PR) angle distributions of (a) equal-width ring approximation, (c) approximated structure using small disks of diameter 0.8 mm, and (e) final simplified pattern composed of one central disk of diameter 2 mm and 174 disks of diameter 0.8 mm. (b, d, f) Corresponding color maps of the simulated beam propagation; intensities are normalized.

**Supplementary Note 3: Simulated Magnetic Field Distribution**

The external magnetic field applied from the Helmholtz coil, $\mu_0 H_{ex}$, was set to 53 mT, corresponding to the minimum field required to reduce the reflected light intensity to the background level, as verified experimentally in Section 4. Fig. S2a and d show the simulation model and layer thicknesses. Fig. S2b shows the $z$-component of the magnetic field distribution on the garnet film surface at $\mu_0 H_{ex}$ = 0 mT. The domain width directly above the magnets widens toward the outside, while the domain width above the brass plate narrows toward the outside. Fig. S2c shows the corresponding distribution at $\mu_0 H_{ex}$ = 53 mT, confirming that the surface magnetic field exceeds the garnet film saturation field over almost the entire aperture, except for a small region directly above the brass fixture near the center. Figs. S2e and S2f show the $z$-component of the magnetic field distribution in the cross-section of the garnet film at $\mu_0 H_{ex}$ = 0 mT and 53 mT, respectively. In both cases, the magnetic field is nearly uniform through the film thickness, confirming that the size and field strength of the permanent magnets are sufficient.

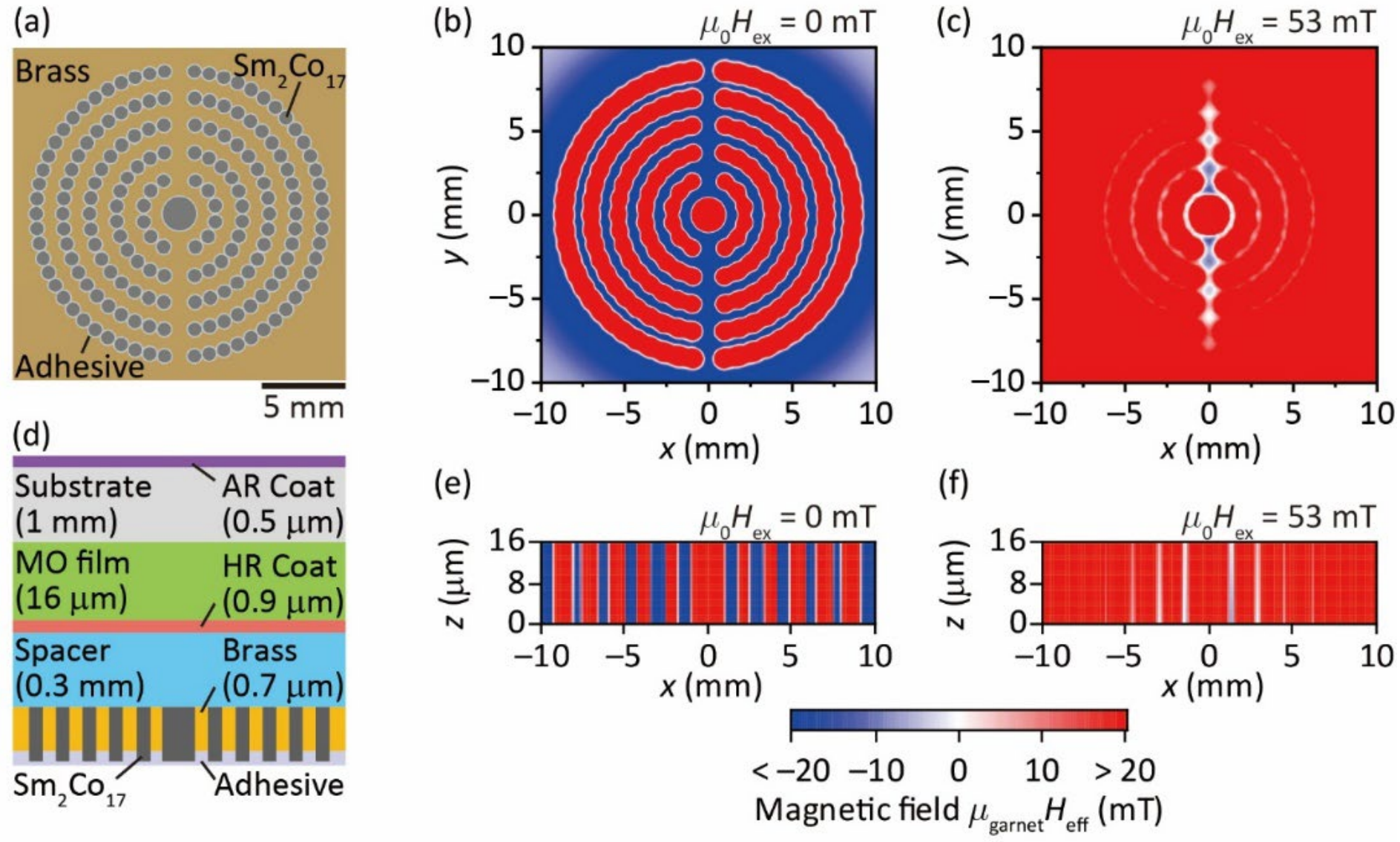

**FIGURE S2.** Magnetic field distribution formed by annularly arranged magnets, with and without an applied external magnetic field. (a, d) The simulation model and the thickness of each layer. Magnetic field distribution at the surface of garnet film at (b) $\mu_0 H_{ex}$ = 0 mT, and (c) 53 mT. (e, f) Cross-sectional magnetic field distribution of $y$ = 0 for (b) and (c).

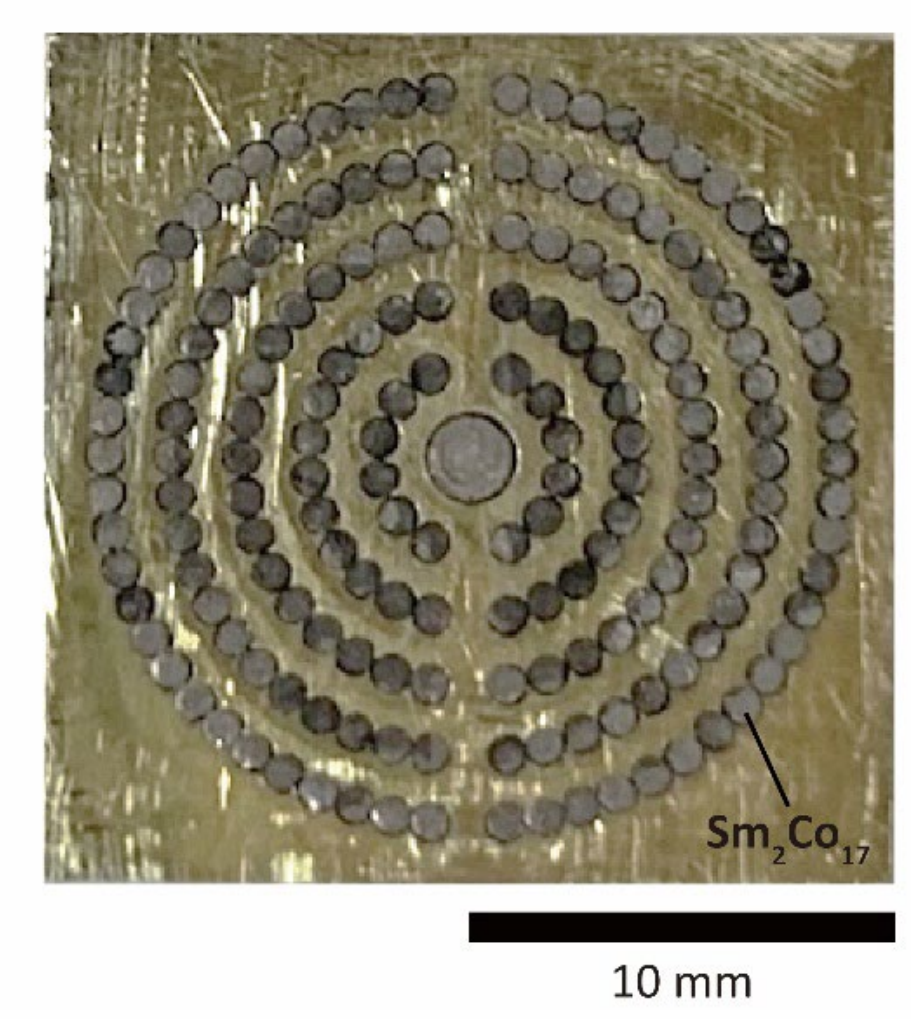

**FIGURE S3.** Photograph showing the overall appearance of the fabricated permanent magnet module.

**TABLE S1.** Material parameters and layer thicknesses used in the magnetic field simulation.

| Layer | Material | Thickness |
|---|---|---|
| AR coating | $SiO_2/Ta_2O_5/SiO_2$ | 0.5 µm |
| Substrate | GGG | 1.0 mm |
| MO film | Magnetic garnet | 16 µm |
| HR coating | $SiO_2/Ta_2O_5$ | 0.9 µm |
| Spacer | Paper | 0.3 mm |
| Brass plate | C2600 | 0.7 mm |
| Magnet | $Sm_2Co_{17}$ | 1.0 mm |
| Adhesive | Cyanoacrylate | 0.4 mm |

**Supplementary Note 4: Magnetic Characterization of Garnet Film**

Magnetic properties were measured with a VSM at room temperature on an uncoated sample, with the paramagnetic background subtracted. An ionizing fan was used to suppress electrostatic noise. A magnetic field of ±1.2 T was applied. The saturation magnetization $M_s$ was 18.9 kA/m, the OP saturation magnetic field $H_{s,OP}$ was 13.9±0.5 mT, and the IP saturation magnetic field $H_{s,IP}$ was 115±0.2 mT.